\documentclass[]{aa}
\usepackage{graphicx}
\usepackage[dvipsnames]{xcolor}
\usepackage[varg]{txfonts}
\usepackage{subcaption}
\usepackage{natbib}
\usepackage{color}
\usepackage{overpic}
\usepackage{pict2e}
\usepackage[colorlinks = true, urlcolor = NavyBlue, linkcolor = NavyBlue, citecolor = NavyBlue, final]{hyperref}
\usepackage{booktabs}

\newcommand{\term}[3]{~$^{#1}\mathrm{{#2}^{#3}}$}

\newcommand{\Eta}{\mathrm{H}}
\newcommand{\Iota}{\mathrm{I}}

\begin{document}

 \title{
 Excitation and charge transfer in low-energy hydrogen atom collisions with neutral manganese and titanium
 \thanks{The complete set of data is made available in electronic form at: {\href{https://github.com/barklem/public-data}{\tt https://github.com/barklem/public-data}}.}
 }

 \titlerunning{Excitation and charge transfer in low-energy H collisions with Mn and Ti}

 \author{J. Grumer \and P. S. Barklem}
 \institute{Theoretical Astrophysics, Department of Physics and Astronomy, Uppsala University, Box 516, SE-751 20 Uppsala, Sweden}

 \date{Received 31 Dec 2019 /
 Accepted 8 March 2020}


\abstract{
Data for inelastic processes due to hydrogen atom collisions with manganese and titanium are needed for accurate modeling of the corresponding spectra in late-type stars.
In this work excitation and charge transfer in low-energy Mn+H and Ti+H collisions have been studied theoretically using a method based on an asymptotic two-electron linear combination of an atomic orbitals model of ionic-covalent interactions in the neutral atom-hydrogen-atom system, together with the multichannel Landau-Zener model to treat the dynamics.
Extensive calculations of charge transfer (mutual neutralization, ion-pair  production), excitation and de-excitation processes in the two collisional  systems are carried out for all transitions between covalent states dissociating to energies below the first ionic limit and the dominating ionic states. Rate coefficients are determined for temperatures in the range 1000 - 20 000 K in steps of 1000 K.
Like for earlier studies of other atomic species, charge transfer processes are found to lead to much larger rate coefficients than excitation processes.
}


\maketitle

\section{Introduction}
\label{sec:intro}
This paper is a continuation of work carried out to provide data for excitation and charge transfer in collisions of hydrogen atoms with astrophysically important elements at temperatures corresponding to photospheres of late-type (FGK) stars.  The ultimate goal is to enable reliable nonlocal thermodynamic equilibrium (NLTE) modeling of the spectra of such stars, and thus accurate determination of stellar properties, including elemental abundances. Since the work of \cite{steenbock_statistical_1984} showing the likely importance of hydrogen collisions on Li abundances in metal-poor stars, collision processes due to hydrogen atoms have been a major uncertainty in abundance analysis of FGK stars \citep[e.g.,][]{Asplund2005,barklem_accurate_2016}.  The aim of this and previous work is to provide calculations based as far as possible on quantum mechanics, to replace the widely used classical Thomson model approach of the Drawin formula \citep{Drawin1968,Drawin1969,steenbock_statistical_1984}.

Work on quantum mechanical calculations of low-energy hydrogen atom collision processes on neutral atoms initially focused on simple targets with a small number of active electrons using quantum chemistry structure calculations and full quantum scattering calculations \citep{Belyaev1999, Belyaev2003, Barklem2003b, Belyaev2010, Barklem2010, Guitou2011, Belyaev2012, Barklem2012}.  More recently, in order to be able to treat more complex elements such as the astrophysically important open-shell elements, (e.g., light open p-shell elements such as C, N, and O, and iron-peak elements with open d-shells) simpler, yet widely applicable, asymptotic methods were developed \citep{Belyaev2013,barklem_excitation_2016,barklem_erratum:_2017}.  These methods have been shown to provide reasonable estimates of the largest, and thus likely the most important, processes.  Recent work by one of us and collaborators has provided data for C and N \citep{Amarsi2019}, O \citep{BarklemExcitationchargetransfer2018a}, and Fe \citep{BarklemExcitationchargetransfer2018}. The calculations generally show\ the importance of charge transfer processes, not considered within the classical model on which the Drawin formula is based, and that the importance of excitation processes is usually significantly overestimated by the Drawin formula. To account for this uncertainty, a scaling factor is often applied and adjusted to match observations.
Applications of quantum mechanical data to modeling stellar spectra have shown that hydrogen collision processes affect modeled spectra and derived abundances \citep[e.g.,][]{Lind2009b,Lind2011,osorio_mg_2015,AmarsiInelasticcollisions7772018,OsorioCalineformation2019, ReggianiNonLTEanalysislatetype2019b,Amarsi3DnonLTEline2019}.

In this work, we turn to two further iron-peak elements, namely manganese and titanium, both important in late-type stellar spectra.  Manganese is believed to be predominantly synthesized in Type Ia supernova explosions, and the abundance of Mn in metal-poor stars thus provides a probe of the physical conditions in supernovae, as well as providing constraints on the progenitors \citep{Seitenzahl2013}.  Manganese is also suggested to be particularly useful as a chemical tag for distinguishing between a thin disk, thick disk and a halo, and accreted halo components \citep{HawkinsUsingchemicaltagging2015}.  Titanium has a large number of spectral lines of both the neutral and singly ionized species in late-type stars, and thus is a very useful diagnostic of stellar parameters (effective temperature, surface gravity) through excitation and ionization equilibrium, complementary to iron.

\begin{table*}[]
\footnotesize
\center
\caption{\label{tab:input_Mn} MnH molecular states in the asymptotic atomic state representation, i.e. possible scattering channels, and associated input data. See the text for further details.
}
\begin{tabular*}{\textwidth}{l @{\extracolsep{\fill}} rcccrrrclccr}
\toprule
$\mathrm{Term}_A$ & $ L_A$ & $2S_A+1$ & $  n$ & $  l$ & $ E_j^\mathrm{Mn}$ [cm$^{-1}$]& $  E_{lim}$ [cm$^{-1}$] & $ E_j$[cm$^{-1}$] & $N_{eq}$ & $  \mathrm{Term}_c$ & $ L_c$ & $2S_c+1$ & $ G^{S_A L_A}_{S_c L_c}$ \\ \midrule
\multicolumn{13}{l}{\textbf{Covalent states:} Mn(($^{2S_c+1}$ $L_c$) $nl$ $^{2S_A+1}$ $L_A$)+H(1s\term{2}{S}{}) } \\ \midrule
$ a~^6\mathrm{S}$   & $  0$ & $  6$ & $  4$ & $  0$ & $     0$ & $ 59960$ & $     0$ & $  2$ & $ a~^7\mathrm{S}$ &$  0$ & $  7$ & $ 0.764$ \\
$ a~^6\mathrm{S}$   & $  0$ & $  6$ & $  4$ & $  0$ & $     0$ & $ 69433$ & $     0$ & $  2$ & $ a~^5\mathrm{S}$ &$  0$ & $  5$ & $ -0.645$ \\
$ a~^6\mathrm{D}$   & $  2$ & $  6$ & $  4$ & $  0$ & $ 17301$ & $ 74546$ & $ 17301$ & $  1$ & $ a~^5\mathrm{D}$ &$  2$ & $  5$ & $ 1.000$ \\
$ z~^8\mathrm{P^o}$ & $  1$ & $  8$ & $  4$ & $  1$ & $ 18572$ & $ 59960$ & $ 18572$ & $  1$ & $ a~^7\mathrm{S}$ &$  0$ & $  7$ & $ 1.000$ \\
$ a~^4\mathrm{D}$   & $  2$ & $  4$ & $  4$ & $  0$ & $ 23509$ & $ 74546$ & $ 23509$ & $  1$ & $ a~^5\mathrm{D}$ &$  2$ & $  5$ & $ 1.000$ \\
$ z~^6\mathrm{P^o}$ & $  1$ & $  6$ & $  4$ & $  1$ & $ 24792$ & $ 59960$ & $ 24792$ & $  1$ & $ a~^7\mathrm{S}$ &$  0$ & $  7$ & $ 1.000$ \\
$ z~^6\mathrm{P^o}$ & $  1$ & $  6$ & $  4$ & $  1$ & $ 24792$ & $ 69433$ & $ 24792$ & $  1$ & $ a~^5\mathrm{S}$ &$  0$ & $  5$ & $ 1.000$ \\
$ a~^4\mathrm{G}$   & $  4$ & $  4$ & $  4$ & $  0$ & $ 25279$ & $ 87531$ & $ 25279$ & $  2$ & $ a~^5\mathrm{G}$ &$  4$ & $  5$ & $ 0.791$ \\
$ a~^4\mathrm{G}$   & $  4$ & $  4$ & $  4$ & $  0$ & $ 25279$ & $ 93175$ & $ 25279$ & $  2$ & $ a~^3\mathrm{G}$ &$  4$ & $  3$ & $ -0.612$ \\
$ a~^4\mathrm{G}$   & $  1$ & $  4$ & $  4$ & $  0$ & $ 27230$ & $ 87531$ & $ 27230$ & $  2$ & $ a~^5\mathrm{G}$ &$  4$ & $  5$ & $ 0.791$ \\
$ a~^4\mathrm{G}$   & $  1$ & $  4$ & $  4$ & $  0$ & $ 27230$ & $ 93175$ & $ 27230$ & $  2$ & $ a~^3\mathrm{G}$ &$  4$ & $  3$ & $ -0.612$ \\
$ b~^4\mathrm{D}$   & $  2$ & $  4$ & $  4$ & $  0$ & $ 30394$ & $ 92789$ & $ 30394$ & $  2$ & $ b~^5\mathrm{D}$ &$  2$ & $  5$ & $ 0.791$ \\
$ b~^4\mathrm{D}$   & $  2$ & $  4$ & $  4$ & $  0$ & $ 30394$ & $ 99774$ & $ 30394$ & $  2$ & $ b~^3\mathrm{D}$ &$  2$ & $  3$ & $ -0.612$ \\
$ z~^4\mathrm{P^o}$ & $  1$ & $  4$ & $  4$ & $  1$ & $ 31047$ & $ 69433$ & $ 31047$ & $  1$ & $ a~^5\mathrm{S}$ &$  0$ & $  5$ & $ 1.000$ \\
$ b~^4\mathrm{P}$   & $  1$ & $  4$ & $  4$ & $  0$ & $ 34208$ & $ 90229$ & $ 34208$ & $  1$ & $ a~^3\mathrm{P}$ &$  1$ & $  3$ & $ 1.000$ \\
$ a~^4\mathrm{H}$   & $  5$ & $  4$ & $  4$ & $  0$ & $ 34268$ & $ 90610$ & $ 34268$ & $  1$ & $ a~^3\mathrm{H}$ &$  5$ & $  3$ & $ 1.000$ \\
$ a~^4\mathrm{F}$   & $  3$ & $  4$ & $  4$ & $  0$ & $ 35038$ & $ 91583$ & $ 35038$ & $  1$ & $ a~^3\mathrm{F}$ &$  3$ & $  3$ & $ 1.000$ \\
$ y~^6\mathrm{P^o}$ & $  1$ & $  6$ & $  4$ & $  1$ & $ 35737$ & $ 59960$ & $ 35737$ & $  1$ & $ a~^7\mathrm{S}$ &$  0$ & $  7$ & $ 1.000$ \\
$ y~^6\mathrm{P^o}$ & $  1$ & $  6$ & $  4$ & $  1$ & $ 35737$ & $ 69433$ & $ 35737$ & $  1$ & $ a~^5\mathrm{S}$ &$  0$ & $  5$ & $ 1.000$ \\
$ a~^2\mathrm{I}$   & $  6$ & $  2$ & $  4$ & $  0$ & $ 37157$ & $101151$ & $ 37157$ & $  2$ & $ a~^3\mathrm{I}$ &$  6$ & $  3$ & $ 0.866$ \\
\multicolumn{13}{c}{$\cdots$ \quad $\cdots$ \quad $\cdots$} \\
$ z~^8\mathrm{F^o}$ & $  3$ & $  8$ & $  4$ & $  3$ & $ 52975$ & $ 59960$ & $ 52975$ & $  1$ & $ a~^7\mathrm{S}$ &$  0$ & $  7$ & $ 1.000$ \\
$ w~^6\mathrm{F^o}$ & $  3$ & $  6$ & $  4$ & $  3$ & $ 52978$ & $ 59960$ & $ 52978$ & $  1$ & $ a~^7\mathrm{S}$ &$  0$ & $  7$ & $ 1.000$ \\
$ y~^4\mathrm{D^o}$ & $  2$ & $  4$ & $  4$ & $  1$ & $ 53133$ & $ 89872$ & $ 53133$ & $  1$ & $ a~^5\mathrm{P}$ &$  1$ & $  5$ & $ 1.000$ \\
$ y~^4\mathrm{D^o}$ & $  2$ & $  4$ & $  4$ & $  1$ & $ 53133$ & $ 96282$ & $ 53133$ & $  1$ & $ b~^3\mathrm{P}$ &$  1$ & $  3$ & $ 1.000$ \\
$ t~^8\mathrm{P^o}$ & $  1$ & $  6$ & $  6$ & $  1$ & $ 53282$ & $ 59960$ & $ 53282$ & $  1$ & $ a~^7\mathrm{S}$ &$  0$ & $  7$ & $ 1.000$ \\
\multicolumn{13}{c}{\ } \\
\multicolumn{5}{l}{\textbf{Ionic states:} {$\mathrm{Mn}^+ (^{2S_A+1} L_A) +\mathrm{H}^-$}(1s$^2$\term{1}{S}{}) } &
$ E_j^{\mathrm{Mn}^+}$ [cm$^{-1}$]&   & $ E_j$[cm$^{-1}$] &\multicolumn{5}{c}{}\\ \midrule
$ a~^7\mathrm{S}$ & $ 0$ & $ 7$ & - & - & $  59960$ & - & $ 53877$ & \multicolumn{5}{c}{} \\
$ a~^5\mathrm{S}$ & $ 0$ & $ 5$ & - & - & $  69433$ & - & $ 63350$ & \multicolumn{5}{c}{} \\
$ a~^5\mathrm{D}$ & $ 2$ & $ 5$ & - & - & $  74546$ & - & $ 68463$ & \multicolumn{5}{c}{} \\
$ a~^5\mathrm{G}$ & $ 4$ & $ 5$ & - & - & $  87531$ & - & $ 81448$ & \multicolumn{5}{c}{} \\
$ a~^3\mathrm{P}$ & $ 1$ & $ 3$ & - & - & $  90229$ & - & $ 84146$ & \multicolumn{5}{c}{} \\
$ a~^5\mathrm{P}$ & $ 1$ & $ 5$ & - & - & $  89872$ & - & $ 83789$ & \multicolumn{5}{c}{} \\
$ a~^3\mathrm{H}$ & $ 5$ & $ 3$ & - & - & $  90610$ & - & $ 84527$ & \multicolumn{5}{c}{} \\
$ a~^3\mathrm{F}$ & $ 3$ & $ 3$ & - & - & $  91583$ & - & $ 85500$ & \multicolumn{5}{c}{} \\
$ b~^5\mathrm{D}$ & $ 2$ & $ 5$ & - & - & $  92789$ & - & $ 86706$ & \multicolumn{5}{c}{} \\
$ a~^3\mathrm{G}$ & $ 4$ & $ 3$ & - & - & $  93175$ & - & $ 87092$ & \multicolumn{5}{c}{} \\
$ b~^3\mathrm{G}$ & $ 4$ & $ 3$ & - & - & $  94835$ & - & $ 88752$ & \multicolumn{5}{c}{} \\
$ b~^3\mathrm{P}$ & $ 1$ & $ 3$ & - & - & $  96282$ & - & $ 90199$ & \multicolumn{5}{c}{} \\
$ a~^3\mathrm{D}$ & $ 2$ & $ 3$ & - & - & $  97802$ & - & $ 91719$ & \multicolumn{5}{c}{} \\
$z~^7\mathrm{P^o}$& $ 1$ & $ 7$ & - & - & $  98574$ & - & $ 92491$ & \multicolumn{5}{c}{} \\
$ a~^1\mathrm{I}$ & $ 6$ & $ 1$ & - & - & $  98680$ & - & $ 92597$ & \multicolumn{5}{c}{} \\
$ a~^1\mathrm{G}$ & $ 4$ & $ 1$ & - & - & $  98862$ & - & $ 92779$ & \multicolumn{5}{c}{} \\
$ b~^3\mathrm{D}$ & $ 2$ & $ 3$ & - & - & $  99774$ & - & $ 93691$ & \multicolumn{5}{c}{} \\
$ a~^3\mathrm{I}$ & $ 6$ & $ 3$ & - & - & $ 101151$ & - & $ 95068$ & \multicolumn{5}{c}{} \\
$ b~^1\mathrm{I}$ & $ 6$ & $ 1$ & - & - & $ 104275$ & - & $ 98192$ & \multicolumn{5}{c}{} \\
$ c~^3\mathrm{P}$ & $ 1$ & $ 3$ & - & - & $ 113266$ & - & $ 107183$ & \multicolumn{5}{c}{} \\
$ e~^3\mathrm{F}$ & $ 3$ & $ 3$ & - & - & $ 113732$ & - & $ 107649$ & \multicolumn{5}{c}{} \\
\bottomrule
\end{tabular*}
\end{table*}

\begin{table*}[]
\footnotesize
\center
\caption{\label{tab:input_Ti} TiH molecular states in the asymptotic atomic state representation, i.e. possible scattering channels, and associated input data. See the text for further details.
}
\begin{tabular*}{\textwidth}{l @{\extracolsep{\fill}} rcccrrrclccr}
\toprule
$\mathrm{Term}_A$ & $ L_A$ & $2S_A+1$ & $  n$ & $  l$ & $ E_j^{\mathrm{Ti}}$ [cm$^{-1}$]& $  E_{lim}$ [cm$^{-1}$] & $ E_j$[cm$^{-1}$] & $N_{eq}$ & $  \mathrm{Term}_c$ & $ L_c$ & $2S_c+1$ & $ G^{S_A L_A}_{S_c L_c}$ \\ \midrule
\multicolumn{13}{l}{\textbf{Covalent states:} $\mathrm{Ti}$(($^{2S_c+1}$ $L_c$) $nl$ $^{2S_A+1}$ $L_A$) $+\mathrm{H}$(1s\term{2}{S}{}) } \\ \midrule
$ a~^3\mathrm{F}$         & $ 3$   & $ 3$   & $ 4$   & $ 0$   & $   0$      & $  59629$   & $   0$      & $ 2$   & $ a~^2\mathrm{F}$   & $ 3$   & $ 2$   & $ -0.577$\\
$ a~^3\mathrm{F}$         & $ 3$   & $ 3$   & $ 4$   & $ 0$   & $   0$      & $  55072$   & $   0$      & $ 2$   & $ a~^4\mathrm{F}$   & $ 3$   & $ 4$   & $ 0.816$ \\
$ a~^5\mathrm{F}$         & $ 3$   & $ 5$   & $ 4$   & $ 0$   & $  6498$    & $  55932$   & $  6498$    & $ 1$   & $ b~^4\mathrm{F}$   & $ 3$   & $ 4$   & $ 1.000$ \\
$ a~^1\mathrm{D}$         & $ 2$   & $ 1$   & $ 4$   & $ 0$   & $  7032$    & $  63578$   & $  7032$    & $ 2$   & $ a~^2\mathrm{D}$   & $ 2$   & $ 2$   & $ 1.000$ \\
$ a~^3\mathrm{P}$         & $ 1$   & $ 3$   & $ 4$   & $ 0$   & $  8325$    & $  71436$   & $  8325$    & $ 2$   & $ b~^2\mathrm{P}$   & $ 1$   & $ 2$   & $ -0.577$\\
$ a~^3\mathrm{P}$         & $ 1$   & $ 3$   & $ 4$   & $ 0$   & $  8325$    & $  64815$   & $  8325$    & $ 2$   & $ b~^4\mathrm{P}$   & $ 1$   & $ 4$   & $ 0.816$ \\
$ b~^3\mathrm{F}$         & $ 3$   & $ 3$   & $ 4$   & $ 0$   & $  11450$   & $  55932$   & $  11450$   & $ 1$   & $ b~^4\mathrm{F}$   & $ 3$   & $ 4$   & $ 1.000$ \\
$ a~^1\mathrm{G}$         & $ 4$   & $ 1$   & $ 4$   & $ 0$   & $  11896$   & $  70108$   & $  11896$   & $ 2$   & $ b~^2\mathrm{G}$   & $ 4$   & $ 2$   & $ 1.000$ \\
$ a~^5\mathrm{P}$         & $ 1$   & $ 5$   & $ 4$   & $ 0$   & $  13833$   & $  64298$   & $  13833$   & $ 1$   & $ a~^4\mathrm{P}$   & $ 1$   & $ 4$   & $ 1.000$ \\
$ a~^3\mathrm{G}$         & $ 4$   & $ 3$   & $ 4$   & $ 0$   & $  14948$   & $  63912$   & $  14948$   & $ 1$   & $ a~^2\mathrm{G}$   & $ 4$   & $ 2$   & $ 1.000$ \\
$ z~^5\mathrm{G}^o$       & $ 4$   & $ 5$   & $ 4$   & $ 1$   & $  15979$   & $  55072$   & $  15979$   & $ 1$   & $ a~^4\mathrm{F}$   & $ 3$   & $ 4$   & $ 1.000$ \\
$ z~^5\mathrm{F}^o$       & $ 3$   & $ 5$   & $ 4$   & $ 1$   & $  16823$   & $  55072$   & $  16823$   & $ 1$   & $ a~^4\mathrm{F}$   & $ 3$   & $ 4$   & $ 1.000$ \\
$ a~^3\mathrm{D}$         & $ 2$   & $ 3$   & $ 4$   & $ 0$   & $  17245$   & $  67553$   & $  17245$   & $ 1$   & $ b~^2\mathrm{D}2$  & $ 2$   & $ 2$   & $ 1.000$ \\
$ b~^3\mathrm{P}$         & $ 1$   & $ 3$   & $ 4$   & $ 0$   & $  17878$   & $  64781$   & $  17878$   & $ 1$   & $ a~^2\mathrm{P}$   & $ 1$   & $ 2$   & $ 1.000$ \\
$ a~^3\mathrm{H}$         & $ 5$   & $ 3$   & $ 4$   & $ 0$   & $  17911$   & $  67577$   & $  17911$   & $ 1$   & $ a~^2\mathrm{H}$   & $ 5$   & $ 2$   & $ 1.000$ \\
$ b~^1\mathrm{G}$         & $ 4$   & $ 1$   & $ 4$   & $ 0$   & $  18065$   & $  63912$   & $  18065$   & $ 1$   & $ a~^2\mathrm{G}$   & $ 4$   & $ 2$   & $ 1.000$ \\
$ z~^5\mathrm{D}^o$       & $ 2$   & $ 5$   & $ 4$   & $ 1$   & $  18375$   & $  55072$   & $  18375$   & $ 1$   & $ a~^4\mathrm{F}$   & $ 3$   & $ 4$   & $ 1.000$ \\
$ c~^3\mathrm{P}$         & $ 1$   & $ 3$   & $ 4$   & $ 0$   & $  18650$   & $  64298$   & $  18650$   & $ 1$   & $ a~^4\mathrm{P}$   & $ 1$   & $ 4$   & $ 1.000$ \\
$ z~^3\mathrm{F}^o$       & $ 3$   & $ 3$   & $ 4$   & $ 1$   & $  19241$   & $  59629$   & $  19241$   & $ 1$   & $ a~^2\mathrm{F}$   & $ 3$   & $ 2$   & $ 1.000$ \\
$ z~^3\mathrm{F}^o$       & $ 3$   & $ 3$   & $ 4$   & $ 1$   & $  19241$   & $  55072$   & $  19241$   & $ 1$   & $ a~^4\mathrm{F}$   & $ 3$   & $ 4$   & $ 1.000$ \\
\multicolumn{13}{c}{$\cdots$ \quad $\cdots$ \quad $\cdots$}\\
$  u~^1\mathrm{G}^o$       & $ 4$   & $ 1$   & $ 5$   & $ 1$   & $  46035$   & $  59529$   & $  46035$   & $ 1$   & $ a~^2\mathrm{F}$   & $ 3$   & $ 2$   & $ 1.000$ \\
$ ~~~^1\mathrm{P}^o$      & $ 1$   & $ 1$   & $ 4$   & $ 1$   & $  46138$   & $  64781$   & $  46138$   & $ 1$   & $ a~^2\mathrm{P}$   & $ 1$   & $ 2$   & $ 1.000$ \\
$ ~~~^1\mathrm{D}$        & $ 2$   & $ 1$   & $ 5$   & $ 0$   & $  46390$   & $  63578$   & $  46390$   & $ 1$   & $ a~^2\mathrm{D}$   & $ 2$   & $ 2$   & $ 1.000$ \\
$  f~^1\mathrm{F}$         & $ 3$   & $ 1$   & $ 4$   & $ 2$   & $  46427$   & $  59629$   & $  46427$   & $ 1$   & $ a~^2\mathrm{F}$   & $ 3$   & $ 2$   & $ 1.000$ \\
$ ~~~^3\mathrm{G}^o$      & $ 4$   & $ 3$   & $ 4$   & $ 1$   & $  46661$   & $  75764$   & $  46661$   & $ 1$   & $ b~^2\mathrm{F}$   & $ 3$   & $ 2$   & $ 1.000$ \\
\multicolumn{13}{c}{\ } \\
\multicolumn{5}{l}{\textbf{Ionic states:} {$\mathrm{Ti}^+ (^{2S_A+1} L_A) +\mathrm{H}^-$}(1s$^2$\term{1}{S}{}) } &
$ E_j^{\mathrm{Ti}^+}$ [cm$^{-1}$]&   & $ E_j$[cm$^{-1}$] & \multicolumn{5}{c}{} \\ \midrule
$   a~^4\mathrm{F}$ & $ 3$ & $ 4$ & - & - &  $  55072$ & - & $  48989$ & \multicolumn{5}{c}{} \\
$   b~^4\mathrm{F}$ & $ 3$ & $ 4$ & - & - &  $  55932$ & - & $  49849$ & \multicolumn{5}{c}{} \\
$   a~^2\mathrm{F}$ & $ 3$ & $ 2$ & - & - &  $  59629$ & - & $  53546$ & \multicolumn{5}{c}{} \\
$   a~^2\mathrm{D}$ & $ 2$ & $ 2$ & - & - &  $  63578$ & - & $  57495$ & \multicolumn{5}{c}{} \\
$   a~^2\mathrm{G}$ & $ 4$ & $ 2$ & - & - &  $  63912$ & - & $  57829$ & \multicolumn{5}{c}{} \\
$   a~^4\mathrm{P}$ & $ 1$ & $ 4$ & - & - &  $  64298$ & - & $  58215$ & \multicolumn{5}{c}{} \\
$   a~^2\mathrm{P}$ & $ 1$ & $ 2$ & - & - &  $  64781$ & - & $  58698$ & \multicolumn{5}{c}{} \\
$   b~^4\mathrm{P}$ & $ 1$ & $ 4$ & - & - &  $  64815$ & - & $  58732$ & \multicolumn{5}{c}{} \\
$  b~^2\mathrm{D}2$ & $ 2$ & $ 2$ & - & - &  $  67553$ & - & $  61470$ & \multicolumn{5}{c}{} \\
$   a~^2\mathrm{H}$ & $ 5$ & $ 2$ & - & - &  $  67577$ & - & $  61494$ & \multicolumn{5}{c}{} \\
$   b~^2\mathrm{G}$ & $ 4$ & $ 2$ & - & - &  $  70108$ & - & $  64025$ & \multicolumn{5}{c}{} \\
$   b~^2\mathrm{P}$ & $ 1$ & $ 2$ & - & - &  $  71436$ & - & $  65353$ & \multicolumn{5}{c}{} \\
$   b~^2\mathrm{F}$ & $ 3$ & $ 2$ & - & - &  $  75764$ & - & $  69681$ & \multicolumn{5}{c}{} \\
$   c~^2\mathrm{D}$ & $ 2$ & $ 2$ & - & - &  $  79947$ & - & $  73864$ & \multicolumn{5}{c}{} \\
$  d~^2\mathrm{D}1$ & $ 2$ & $ 2$ & - & - &  $  87157$ & - & $  81074$ & \multicolumn{5}{c}{} \\
\bottomrule
\end{tabular*}
\end{table*}

Work on the spectra of Mn and Ti in nonlocal thermodynamic equilibrium (NLTE) was carried out by Bergemann and collaborators. In the case of Mn, early studies \citep{BergemannFormationMnlines2007, BergemannNLTEabundancesMn2008} employed the Drawin formula to model the effects of collisions with hydrogen atoms, with a scaling factor of 0.05, to reflect that the evidence at the time suggested the Drawin formula strongly overestimated the collisional excitation rates. The more recent work of \cite{bergemann_observational_2019} uses quantum mechanical calculations of hydrogen collision processes for Mn by \cite{belyaev_atomic_2017} (hereafter BV17), and thus includes excitation and charge transfer processes. These collision calculations were made with an asymptotic model approach employing semiempirical couplings for the dominant interaction and the multichannel Landau-Zener model for the dynamics, and considers the dominant molecular symmetry ($^7\Sigma^+$).
Work on the Ti spectrum in NLTE by \cite{bergemann_ionization_2011} also adopted the Drawin formula, but with a much larger scaling factor.
The abundance of lines in the solar spectrum allowed Bergemann to attempt to constrain the efficiency of hydrogen collisions by minimizing the abundance scatter, and they found a scaling factor of~3.
No quantum mechanical calculations have yet been done for Ti.

In this paper, we present the results of calculations for hydrogen collision processes on Mn and Ti using an asymptotic model approach based on a theoretical two-electron linear combination of atomic orbitals (LCAO) asymptotic model for the atomic interactions, together with the multi-channel Landau-Zener model of nonadiabatic collision dynamics.  This same method was used earlier for a number of atoms mentioned above.  In the case of Mn, our calculations provide an independent check on the work of BV17 (which uses a different method for determining the couplings - semiempirical vs. asymptotic), and extend to additional excited molecular states beyond the $^7\Sigma^+$ symmetry. In the case of Ti, our calculations provide data that can be used in modeling with a sounder physical basis than that of the Drawin formula.

\begin{table*}[]
\footnotesize
\center
\caption{\label{tab:states_Mn} Possible MnH molecular symmetries (terms) for the asymptotic states included in the calculations, where asymptotic states with different Mn cores have been merged. See the text for further details.}
\begin{tabular*}{\textwidth}{llll@{\extracolsep{\fill}}}
\toprule
Index & Term$_A$ & $g_{total}$ & Molecular terms \\ \midrule
\multicolumn{4}{l}{\textbf{Covalent states:} Mn($^{2S_A+1}$ $L_A$)+H(1s\term{2}{S}{}) } \\ \midrule
$  1$ & $ a~^6\mathrm{S}$   &   12& $   ^{5}\Sigma^+,\ ^{7}\Sigma^+$ \\
$  2$ & $ a~^6\mathrm{D}$   &   60& $   ^{5}\Sigma^+,\ ^{5}\Pi,\     ^{5}\Delta,\   ^{7}\Sigma^+,\ ^{7}\Pi,\     ^{7}\Delta$ \\
$  3$ & $ z~^8\mathrm{P}^o$ &   48& $   ^{7}\Sigma^+,\ ^{7}\Pi,\   ^{9}\Sigma^+,\ ^{9}\Pi$ \\
$  4$ & $ a~^4\mathrm{D}$   &   40& $   ^{3}\Sigma^+,\ ^{3}\Pi,\     ^{3}\Delta,\   ^{5}\Sigma^+,\ ^{5}\Pi,\     ^{5}\Delta$ \\
$  5$ & $ z~^6\mathrm{P}^o$ &   36& $   ^{5}\Sigma^+,\ ^{5}\Pi,\   ^{7}\Sigma^+,\ ^{7}\Pi$ \\
$  6$ & $ a~^4\mathrm{G}$   &   72& $   ^{3}\Sigma^+,\ ^{3}\Pi,\     ^{3}\Delta,\ ^{3}\Phi,\     ^{3}\Gamma,\   ^{5}\Sigma^+,\ ^{5}\Pi,\     ^{5}\Delta,\ ^{5}\Phi,\     ^{5}\Gamma$ \\
$  7$ & $ a~^4\mathrm{G}$   &   24& $   ^{3}\Sigma^-,\ ^{3}\Pi,\   ^{5}\Sigma^-,\ ^{5}\Pi$ \\
$  8$ & $ b~^4\mathrm{D}$   &   40& $   ^{3}\Sigma^+,\ ^{3}\Pi,\     ^{3}\Delta,\   ^{5}\Sigma^+,\ ^{5}\Pi,\     ^{5}\Delta$ \\
$  9$ & $ z~^4\mathrm{P}^o$ &   24& $   ^{3}\Sigma^+,\ ^{3}\Pi,\   ^{5}\Sigma^+,\ ^{5}\Pi$ \\
$ 10$ & $ b~^4\mathrm{P}$   &   24& $   ^{3}\Sigma^-,\ ^{3}\Pi,\   ^{5}\Sigma^-,\ ^{5}\Pi$ \\
$ 11$ & $ a~^4\mathrm{H}$   &   88& $   ^{3}\Sigma^-,\ ^{3}\Pi,\     ^{3}\Delta,\ ^{3}\Phi,\     ^{3}\Gamma,\ ^{3}\Eta,\   ^{5}\Sigma^-,\ ^{5}\Pi,\     ^{5}\Delta,\ ^{5}\Phi,\     ^{5}\Gamma,\ ^{5}\Eta$ \\
$ 12$ & $ a~^4\mathrm{F}$   &   56& $   ^{3}\Sigma^-,\ ^{3}\Pi,\     ^{3}\Delta,\ ^{3}\Phi,\   ^{5}\Sigma^-,\ ^{5}\Pi,\     ^{5}\Delta,\ ^{5}\Phi$ \\
$ 13$ & $ y~^6\mathrm{P}^o$ &   36& $   ^{5}\Sigma^+,\ ^{5}\Pi,\   ^{7}\Sigma^+,\ ^{7}\Pi$ \\
$ 14$ & $ a~^2\mathrm{I}$   &   52& $   ^{1}\Sigma^+,\ ^{1}\Pi,\     ^{1}\Delta,\ ^{1}\Phi,\     ^{1}\Gamma,\ ^{1}\Eta,\      ^{1}\Iota,\   ^{3}\Sigma^+,\ ^{3}\Pi,\     ^{3}\Delta,\ ^{3}\Phi,\     ^{3}\Gamma,\ ^{3}\Eta,\      ^{3}\Iota$ \\
$ 15$ & $ b~^4\mathrm{G}$   &   72& $   ^{3}\Sigma^+,\ ^{3}\Pi,\     ^{3}\Delta,\ ^{3}\Phi,\     ^{3}\Gamma,\   ^{5}\Sigma^+,\ ^{5}\Pi,\     ^{5}\Delta,\ ^{5}\Phi,\     ^{5}\Gamma$ \\
$ 16$ & $ a~^2\mathrm{P}$   &   12& $   ^{1}\Sigma^-,\ ^{1}\Pi,\   ^{3}\Sigma^-,\ ^{3}\Pi$ \\
$ 17$ & $ a~^2\mathrm{H}$   &   44& $   ^{1}\Sigma^-,\ ^{1}\Pi,\     ^{1}\Delta,\ ^{1}\Phi,\     ^{1}\Gamma,\ ^{1}\Eta,\   ^{3}\Sigma^-,\ ^{3}\Pi,\     ^{3}\Delta,\ ^{3}\Phi,\     ^{3}\Gamma,\ ^{3}\Eta$ \\
$ 18$ & $ a~^2\mathrm{F}$   &   28& $   ^{1}\Sigma^-,\ ^{1}\Pi,\     ^{1}\Delta,\ ^{1}\Phi,\   ^{3}\Sigma^-,\ ^{3}\Pi,\     ^{3}\Delta,\ ^{3}\Phi$ \\
$ 19$ & $ e~^8\mathrm{S}$   &   16& $   ^{7}\Sigma^+,\   ^{9}\Sigma^+$ \\
$ 20$ & $ a~^2\mathrm{G}$   &   36& $   ^{1}\Sigma^+,\ ^{1}\Pi,\     ^{1}\Delta,\ ^{1}\Phi,\     ^{1}\Gamma,\   ^{3}\Sigma^+,\ ^{3}\Pi,\     ^{3}\Delta,\ ^{3}\Phi,\     ^{3}\Gamma$ \\
$\cdots$ & $\cdots$ & $\cdots$ & \quad $\cdots$\\
$ 50$ & $ x~^6\mathrm{D}^o$ &   60& $   ^{5}\Sigma^-,\ ^{5}\Pi,\     ^{5}\Delta,\   ^{7}\Sigma^-,\ ^{7}\Pi,\     ^{7}\Delta$ \\
$ 51$ & $ z~^8\mathrm{F}^o$ &  112& $   ^{7}\Sigma^+,\ ^{7}\Pi,\     ^{7}\Delta,\ ^{7}\Phi,\   ^{9}\Sigma^+,\ ^{9}\Pi,\     ^{9}\Delta,\ ^{9}\Phi$ \\
$ 52$ & $ w~^6\mathrm{F}^o$ &   84& $   ^{5}\Sigma^+,\ ^{5}\Pi,\     ^{5}\Delta,\ ^{5}\Phi,\   ^{7}\Sigma^+,\ ^{7}\Pi,\     ^{7}\Delta,\ ^{7}\Phi$ \\
$ 53$ & $ y~^4\mathrm{D}^o$ &   40& $   ^{3}\Sigma^-,\ ^{3}\Pi,\     ^{3}\Delta,\   ^{5}\Sigma^-,\ ^{5}\Pi,\     ^{5}\Delta$ \\
$ 54$ & $ t~^8\mathrm{P}^o$ &   36& $   ^{5}\Sigma^+,\ ^{5}\Pi,\   ^{7}\Sigma^+,\ ^{7}\Pi$ \\
\multicolumn{4}{c}{\ } \\
\multicolumn{4}{l}{\textbf{Ionic states:} {$\mathrm{Mn}^+(^{2S_A+1} L_A) +\mathrm{H}^-$}(1s$^2$\term{1}{S}{}) } \\ \midrule
$ 55$ & $ a~^7\mathrm{S}$   &    7& $   ^{7}\Sigma^+$ \\
$ 56$ & $ a~^5\mathrm{S}$   &    5& $   ^{5}\Sigma^+$ \\
$ 57$ & $ a~^5\mathrm{D}$   &   25& $   ^{5}\Sigma^+,\ ^{5}\Pi,\     ^{5}\Delta$ \\
$ 58$ & $ a~^5\mathrm{G}$   &   45& $   ^{5}\Sigma^+,\ ^{5}\Pi,\     ^{5}\Delta,\ ^{5}\Phi,\     ^{5}\Gamma$ \\
$ 59$ & $ a~^3\mathrm{P}$   &    9& $   ^{3}\Sigma^-,\ ^{3}\Pi$ \\
$ 60$ & $ a~^5\mathrm{P}$   &   15& $   ^{5}\Sigma^-,\ ^{5}\Pi$ \\
$ 61$ & $ a~^3\mathrm{H}$   &   33& $   ^{3}\Sigma^-,\ ^{3}\Pi,\     ^{3}\Delta,\ ^{3}\Phi,\     ^{3}\Gamma,\ ^{3}\Eta$ \\
$ 62$ & $ a~^3\mathrm{F}$   &   21& $   ^{3}\Sigma^-,\ ^{3}\Pi,\     ^{3}\Delta,\ ^{3}\Phi$ \\
$ 63$ & $ b~^5\mathrm{D}$   &   25& $   ^{5}\Sigma^+,\ ^{5}\Pi,\     ^{5}\Delta$ \\
$ 64$ & $ a~^3\mathrm{G}$   &   27& $   ^{3}\Sigma^+,\ ^{3}\Pi,\     ^{3}\Delta,\ ^{3}\Phi,\     ^{3}\Gamma$ \\
$ 65$ & $ b~^3\mathrm{G}$   &   27& $   ^{3}\Sigma^+,\ ^{3}\Pi,\     ^{3}\Delta,\ ^{3}\Phi,\     ^{3}\Gamma$ \\
$ 66$ & $ b~^3\mathrm{P}$   &    9& $   ^{3}\Sigma^-,\ ^{3}\Pi$ \\
$ 67$ & $ a~^3\mathrm{D}$   &   15& $   ^{3}\Sigma^+,\ ^{3}\Pi,\     ^{3}\Delta$ \\
$ 68$ & $ z~^7\mathrm{P}^o$ &  21& $   ^{7}\Sigma^+,\ ^{7}\Pi$ \\
$ 69$ & $ a~^1\mathrm{I}$   &   13& $   ^{1}\Sigma^+,\ ^{1}\Pi,\     ^{1}\Delta,\ ^{1}\Phi,\     ^{1}\Gamma,\ ^{1}\Eta,\      ^{1}\Iota$ \\
$ 70$ & $ a~^1\mathrm{G}$   &    9& $   ^{1}\Sigma^+,\ ^{1}\Pi,\     ^{1}\Delta,\ ^{1}\Phi,\     ^{1}\Gamma$ \\
$ 71$ & $ b~^3\mathrm{D}$   &   15& $   ^{3}\Sigma^+,\ ^{3}\Pi,\     ^{3}\Delta$ \\
$ 72$ & $ a~^3\mathrm{I}$   &   39& $   ^{3}\Sigma^+,\ ^{3}\Pi,\     ^{3}\Delta,\ ^{3}\Phi,\     ^{3}\Gamma,\ ^{3}\Eta,\      ^{3}\Iota$ \\
$ 73$ & $ b~^1\mathrm{I}$   &   13& $   ^{1}\Sigma^+,\ ^{1}\Pi,\     ^{1}\Delta,\ ^{1}\Phi,\     ^{1}\Gamma,\ ^{1}\Eta,\      ^{1}\Iota$ \\
$ 74$ & $ c~^3\mathrm{P}$   &    9& $   ^{3}\Sigma^-,\ ^{3}\Pi$ \\
$ 75$ & $ e~^3\mathrm{F}$   &   21& $   ^{3}\Sigma^-,\ ^{3}\Pi,\     ^{3}\Delta,\ ^{3}\Phi$ \\
\end{tabular*}
\begin{subtable}{\textwidth}
    \begin{tabular*}{\textwidth}{rllllllll@{\extracolsep{\fill}}}
    \midrule
    Symmetries to calculate (23): &  $^{7,5,3,1}\Sigma^+$ & $^{5,3}\Sigma^-$ & $^{7,5,3,1}\Pi$ & $^{5,3,1}\Delta$ & $^{5,3,1}\Phi$ & $^{5,3,1}\Gamma$ & $^{3,1}\Eta$ & $^{3,1}\Iota$ \\
                               g: & [7, 5, 3, 1] & [5, 3] &  [14, 10, 6, 2] &  [10, 6, 2] &  [10, 6, 2] &  [10, 6, 2] & [6, 2] & [6, 2] \\ \bottomrule
    \end{tabular*}
\end{subtable}
\end{table*}

\begin{table*}[]
\footnotesize
\center
\caption{\label{tab:states_Ti} Possible TiH molecular symmetries (terms) for each asymptotic state $j$ included in the calculations, where asymptotic states with different Ti cores have been merged. See the text for further details.}
\begin{tabular*}{\textwidth}{llll@{\extracolsep{\fill}}}
\toprule
Index & Term$_A$ & $g_{total}$ & Molecular terms \\ \midrule
\multicolumn{4}{l}{\textbf{Covalent states:} Ti($^{2S_A+1}$ $L_A$)+H(1s\term{2}{S}{}) } \\ \midrule
   1 & $   a~^3\mathrm{F} $ &   42& $   ^{2}\Sigma^-,\  ^{2}\Pi,\     ^{2}\Delta,\ ^{2}\Phi,\   ^{4}\Sigma^-,\  ^{4}\Pi,\     ^{4}\Delta,\ ^{4}\Phi$ \\
   2 & $   a~^5\mathrm{F} $ &   70& $   ^{4}\Sigma^-,\  ^{4}\Pi,\     ^{4}\Delta,\ ^{4}\Phi,\   ^{6}\Sigma^-,\  ^{6}\Pi,\     ^{6}\Delta,\ ^{6}\Phi$ \\
   3 & $   a~^1\mathrm{D} $ &   10& $   ^{2}\Sigma^+,\  ^{2}\Pi,\     ^{2}\Delta$ \\
   4 & $   a~^3\mathrm{P} $ &   18& $   ^{2}\Sigma^-,\  ^{2}\Pi,\   ^{4}\Sigma^-,\  ^{4}\Pi$ \\
   5 & $   b~^3\mathrm{F} $ &   42& $   ^{2}\Sigma^-,\  ^{2}\Pi,\     ^{2}\Delta,\ ^{2}\Phi,\   ^{4}\Sigma^-,\  ^{4}\Pi,\     ^{4}\Delta,\ ^{4}\Phi$ \\
   6 & $   a~^1\mathrm{G} $ &   18& $   ^{2}\Sigma^+,\  ^{2}\Pi,\     ^{2}\Delta,\ ^{2}\Phi,\     ^{2}\Gamma$ \\
   7 & $   a~^5\mathrm{P} $ &   30& $   ^{4}\Sigma^-,\  ^{4}\Pi,\   ^{6}\Sigma^-,\  ^{6}\Pi$ \\
   8 & $   a~^3\mathrm{G} $ &   54& $   ^{2}\Sigma^+,\  ^{2}\Pi,\     ^{2}\Delta,\ ^{2}\Phi,\     ^{2}\Gamma,\   ^{4}\Sigma^+,\  ^{4}\Pi,\     ^{4}\Delta,\ ^{4}\Phi,\     ^{4}\Gamma$ \\
   9 & $ z~^5\mathrm{G}^o $ &   90& $   ^{4}\Sigma^-,\  ^{4}\Pi,\     ^{4}\Delta,\ ^{4}\Phi,\     ^{4}\Gamma,\   ^{6}\Sigma^-,\  ^{6}\Pi,\     ^{6}\Delta,\ ^{6}\Phi,\     ^{6}\Gamma$ \\
  10 & $ z~^5\mathrm{F}^o $ &   70& $   ^{4}\Sigma^+,\  ^{4}\Pi,\     ^{4}\Delta,\ ^{4}\Phi,\   ^{6}\Sigma^+,\  ^{6}\Pi,\     ^{6}\Delta,\ ^{6}\Phi$ \\
  11 & $   a~^3\mathrm{D} $ &   30& $   ^{2}\Sigma^+,\  ^{2}\Pi,\     ^{2}\Delta,\   ^{4}\Sigma^+,\  ^{4}\Pi,\     ^{4}\Delta$ \\
  12 & $   b~^3\mathrm{P} $ &   18& $   ^{2}\Sigma^-,\  ^{2}\Pi,\   ^{4}\Sigma^-,\  ^{4}\Pi$ \\
  13 & $   a~^3\mathrm{H} $ &   66& $   ^{2}\Sigma^-,\  ^{2}\Pi,\     ^{2}\Delta,\ ^{2}\Phi,\     ^{2}\Gamma,\ ^{2}\Eta,\   ^{4}\Sigma^-,\  ^{4}\Pi,\     ^{4}\Delta,\ ^{4}\Phi,\     ^{4}\Gamma,\ ^{4}\Eta$ \\
  14 & $   b~^1\mathrm{G} $ &   18& $   ^{2}\Sigma^+,\  ^{2}\Pi,\     ^{2}\Delta,\ ^{2}\Phi,\     ^{2}\Gamma$ \\
  15 & $ z~^5\mathrm{D}^o $ &   50& $   ^{4}\Sigma^-,\  ^{4}\Pi,\     ^{4}\Delta,\   ^{6}\Sigma^-,\  ^{6}\Pi,\     ^{6}\Delta$ \\
  16 & $   c~^3\mathrm{P} $ &   18& $   ^{2}\Sigma^-,\  ^{2}\Pi,\   ^{4}\Sigma^-,\  ^{4}\Pi$ \\
  17 & $ z~^3\mathrm{F}^o $ &   42& $   ^{2}\Sigma^+,\  ^{2}\Pi,\     ^{2}\Delta,\ ^{2}\Phi,\   ^{4}\Sigma^+,\  ^{4}\Pi,\     ^{4}\Delta,\ ^{4}\Phi$ \\
  18 & $ z~^3\mathrm{D}^o $ &   30& $   ^{2}\Sigma^-,\  ^{2}\Pi,\     ^{2}\Delta,\   ^{4}\Sigma^-,\  ^{4}\Pi,\     ^{4}\Delta$ \\
  19 & $   a~^1\mathrm{P} $ &    6& $   ^{2}\Sigma^-,\  ^{2}\Pi$ \\
  20 & $   b~^1\mathrm{D} $ &   10& $   ^{2}\Sigma^+,\  ^{2}\Pi,\     ^{2}\Delta$ \\
$\cdots$ & $\cdots$ & $\cdots$ & \quad $\cdots$\\
 142 & $ u~^1\mathrm{G}^o $ &   18& $   ^{2}\Sigma^-,\  ^{2}\Pi,\     ^{2}\Delta,\ ^{2}\Phi,\     ^{2}\Gamma$ \\
 143 & $  ~^1\mathrm{P}^o $ &    6& $   ^{2}\Sigma^+,\  ^{2}\Pi$ \\
 144 & $    ~^1\mathrm{D} $ &   10& $   ^{2}\Sigma^+,\  ^{2}\Pi,\     ^{2}\Delta$ \\
 145 & $   f~^1\mathrm{F} $ &   14& $   ^{2}\Sigma^-,\  ^{2}\Pi,\     ^{2}\Delta,\ ^{2}\Phi$ \\
 146 & $  ~^3\mathrm{G}^o $ &   54& $   ^{2}\Sigma^-,\  ^{2}\Pi,\     ^{2}\Delta,\ ^{2}\Phi,\     ^{2}\Gamma,\   ^{4}\Sigma^-,\  ^{4}\Pi,\     ^{4}\Delta,\ ^{4}\Phi,\     ^{4}\Gamma$ \\
\multicolumn{4}{c}{\ } \\
\multicolumn{4}{l}{\textbf{Ionic states:} {$\mathrm{Ti}^+(^{2S_A+1} L_A) +\mathrm{H}^-$}(1s$^2$\term{1}{S}{}) } \\ \midrule
 147 & $  a~^4\mathrm{F}  $ &   28& $   ^{4}\Sigma^-,\  ^{4}\Pi,\     ^{4}\Delta,\ ^{4}\Phi$ \\
 148 & $  b~^4\mathrm{F}  $ &   28& $   ^{4}\Sigma^-,\  ^{4}\Pi,\     ^{4}\Delta,\ ^{4}\Phi$ \\
 149 & $  a~^2\mathrm{F}  $ &   14& $   ^{2}\Sigma^-,\  ^{2}\Pi,\     ^{2}\Delta,\ ^{2}\Phi$ \\
 150 & $  a~^2\mathrm{D}  $ &   10& $   ^{2}\Sigma^+,\  ^{2}\Pi,\     ^{2}\Delta$ \\
 151 & $  a~^2\mathrm{G}  $ &   18& $   ^{2}\Sigma^+,\  ^{2}\Pi,\     ^{2}\Delta,\ ^{2}\Phi,\     ^{2}\Gamma$ \\
 152 & $  a~^4\mathrm{P}  $ &   12& $   ^{4}\Sigma^-,\  ^{4}\Pi$ \\
 153 & $  a~^2\mathrm{P}  $ &    6& $   ^{2}\Sigma^-,\  ^{2}\Pi$ \\
 154 & $  b~^4\mathrm{P}  $ &   12& $   ^{4}\Sigma^-,\  ^{4}\Pi$ \\
 155 & $  b~^2\mathrm{D2} $ &   10& $   ^{2}\Sigma^+,\  ^{2}\Pi,\     ^{2}\Delta$ \\
 156 & $  a~^2\mathrm{H}  $ &   22& $   ^{2}\Sigma^-,\  ^{2}\Pi,\     ^{2}\Delta,\ ^{2}\Phi,\     ^{2}\Gamma,\ ^{2}\Eta$ \\
 157 & $  b~^2\mathrm{G}  $ &   18& $   ^{2}\Sigma^+,\  ^{2}\Pi,\     ^{2}\Delta,\ ^{2}\Phi,\     ^{2}\Gamma$ \\
 158 & $  b~^2\mathrm{P}  $ &    6& $   ^{2}\Sigma^-,\  ^{2}\Pi$ \\
 159 & $  b~^2\mathrm{F}  $ &   14& $   ^{2}\Sigma^-,\  ^{2}\Pi,\     ^{2}\Delta,\ ^{2}\Phi$ \\
 160 & $  c~^2\mathrm{D}  $ &   10& $   ^{2}\Sigma^+,\  ^{2}\Pi,\     ^{2}\Delta$ \\
 161 & $  d~^2\mathrm{D1} $ &   10& $   ^{2}\Sigma^+,\  ^{2}\Pi,\     ^{2}\Delta$ \\
\end{tabular*}

\begin{subtable}{\textwidth}
    \begin{tabular*}{\textwidth}{rlllllll@{\extracolsep{\fill}}}
    \midrule
    Symmetries to calculate (11): & $^{2}\Sigma^+$ & $^{4,2}\Sigma^-$ & $^{4,2}\Pi$ & $^{4,2}\Delta$ & $^{4,2}\Phi$ & $^{2}\Gamma$ & $^{2}\Eta$ \\
                               g: & [2] & [4, 2] &  [8, 4] &  [8, 4] &  [8, 4] &  [4] & [4] \\ \bottomrule
    \end{tabular*}
\end{subtable}
\end{table*}

\section{Calculations}
The theoretical method used in this work is presented in \cite{barklem_excitation_2016, barklem_erratum:_2017} (hereafter B16).  It is based on the LCAO asymptotic model of ionic-covalent interactions \citep{Grice1974, Adelman1977, Anstee1992} and the multichannel Landau-Zener model of the nonadiabatic processes occurring at avoided crossings as formulated by \cite{Belyaev1993, Belyaev2003}.  The B16 paper should be consulted for details of the model, as well as definition of the notation used here.

In \cite{BarklemExcitationchargetransfer2018a}, the possibility to treat covalent states with hydrogen in its $n = 2$ state was implemented in the model. Such channels where hydrogen is excited are important for cases when the two colliding atoms have comparable ionization energies, and thus dissociated states involving H($n=2$) exist below the first ionic limit (i.e., the dissociation limit of state corresponding asymptotically to the ground state of the ion, namely $\mathrm{Mn}^+(a^7S) +\mathrm{H}^-$ and $\mathrm{Ti}^+(a^4F) +\mathrm{H}^-$. The ionization energies of Mn and Ti are 7.43 and 6.83~eV, respectively, resulting in corresponding ionic limits of 6.68 and 6.08~eV. This means covalent states dissociating to Mn/Ti + H($n = 2$) are well above the ionic limit and excited states of H can be excluded.  Similarly, channels involving negative ions of the target ($\mathrm{Mn/Ti}^- +\mathrm{H}^+$) have asymptotic energies well above the ionic limit.

The input data to the calculations have been assembled predominantly based on atomic data obtained from the NIST atomic spectra database (ASD) \citep{NIST_5.7.1}, with sources for the level data being \cite{sugar_corliss_atomic_1985} and \cite{saloman_energy_2012} for Mn and Ti, respectively. The required coefficients of fractional parentage are taken from standard tabulations \cite[e.g.,a][]{sobelman_atomic_1979}, and in the case of mixed configurations are combined using the method of \cite{Kelly1959}, based on standard Racah algebra.

\begin{figure*}
    \centering
    \includegraphics[width=1.0\textwidth]{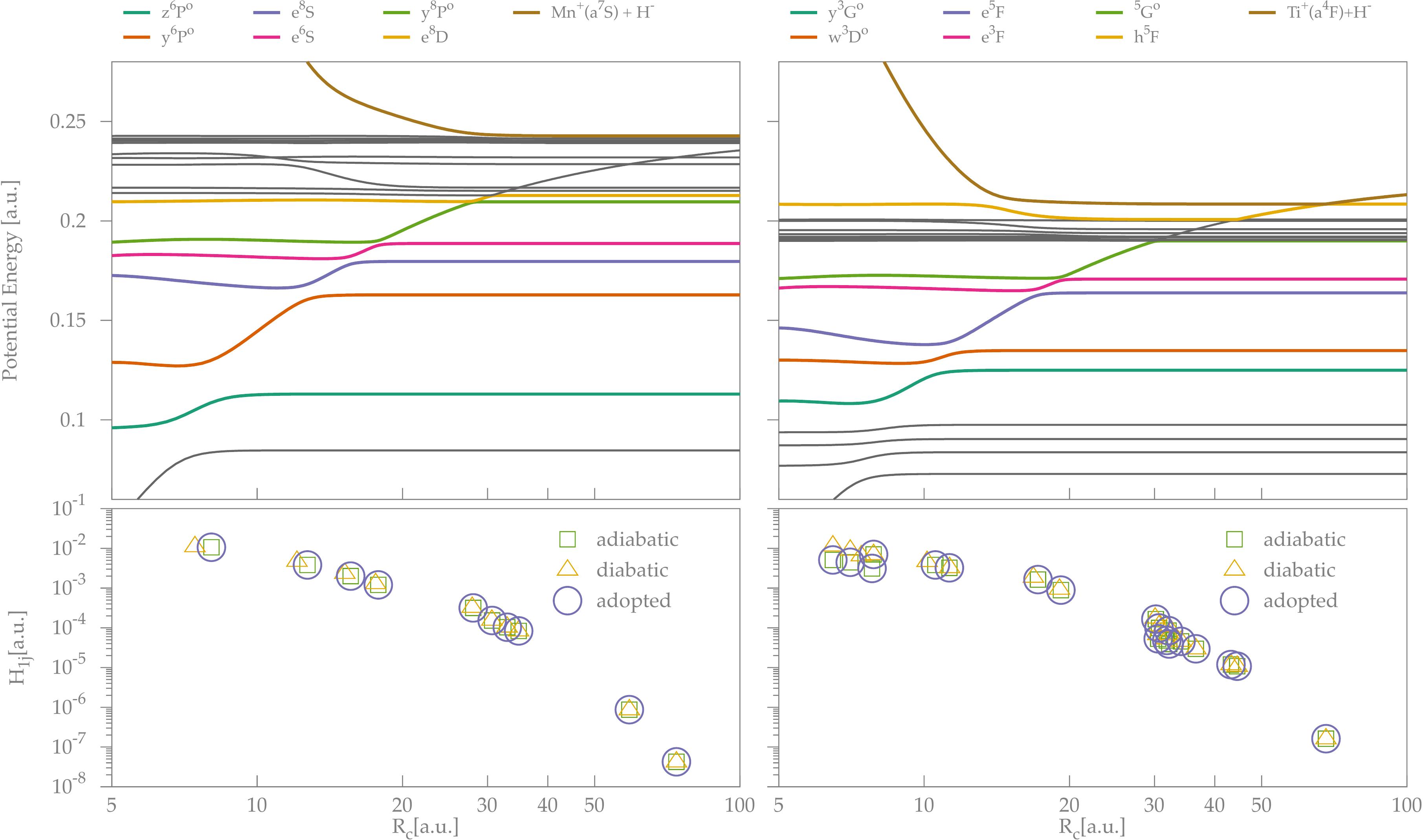}
    \caption{Example potential energies (\textbf{upper}) and couplings (\textbf{lower}) from the LCAO model, for Mn+H with the Mn$^+(a~^7\mathrm{S})$ core in the $^7\Sigma^+$ symmetry (\textbf{left}), and Ti+H with the Ti$^+(a~^4\mathrm{F})$ core in the $^4\Sigma^-$ symmetry (\textbf{right}). A selected subset of states are labeled in the upper panels, and, for the covalent states, only the label of the target Mn or Ti atom is given; i.e. a hydrogen atom in its ground state is implied. Couplings H$_{1 j}$ at avoided crossings are plotted against the crossing distance $R_c$, and results are shown for the adiabatic and diabatic representations, with the values adopted in the circled calculations.}
    \label{fig:pots_couplings}
\end{figure*}

Covalent (Mn/Ti + H) and ionic (Mn$^+$/Ti$^+$ + H$^-$) asymptotic states considered in the calculations are presented Tables \ref{tab:input_Mn} and \ref{tab:input_Ti}. Note that not all states are shown, but the full tables are available electronically (see footnote on the first page for details). The data is essentially presented in line with the notation from the LCAO model in B16, which also contains a detailed description of the method.
Using A to denote either Mn or Ti, the first column in these two tables give the LS term of A or A$^+$, followed by explicit values for the corresponding total orbital angular momentum $L_A$ and spin, $S_A$. The active electron in the interaction is then defined by its principal and angular momentum quantum numbers, $n$ and $l$. The corresponding state energy for A or A$^+$ relative to the ground term of A is given by $E_j^\mathrm{A/{A^+}}$ and the total asymptotic molecular energy is given by $E_j$, with the sum of the ground term energies of A and H as the zero point. In the the covalent case, $E_\mathrm{lim}$ denotes the corresponding series limit, again relative to the ground term energy of A. Next, $N_\mathrm{eq}$ gives the number of equivalent active electrons in the active subshell of A. Finally, Term$_c$, $L_c,$ and $S_c$ define the term, orbital angular momentum, and spin quantum numbers of the core, and $G^{S_A L_A}_{S_c L_c}$ is the coefficient of fractional parentage used in the necessary decoupling of the active electron. For all cases considered here, hydrogen or its anion are always assumed to be in their ground states.

Tables \ref{tab:states_Mn} and \ref{tab:states_Ti} list the possible molecular symmetries resulting from the considered asymptotic scattering channels for Mn and Ti, respectively. Similar to the input data, both covalent and ionic channels are presented, with the difference being that the asymptotic states involving different cores (i.e., with fractional parentage) have been merged into single channels with total statistical weights $g_\mathrm{total}$. At the bottom of the tables, the symmetries for which there are both covalent and ionic states, which thus correspond to covalent-ionic interactions to be calculated, are listed together with their corresponding statistical weights $g$.

In summary, the Mn+H calculation includes 54 covalent states and 21 ionic states, resulting in 23 symmetries that need to be calculated, which includes all covalent states dissociating to energies below the first ionic limit. The Ti+H calculation includes 146 covalent states and 15 ionic states, resulting in 11 symmetries that need to be calculated. Here, all covalent states dissociating to energies below 47000 cm$^{-1}$ were included, very close to the first ionic limit at 48989 cm$^{-1}$; higher lying states either involve highly excited cores or have crossings only at very large internuclear distance ($>200$ atomic units).

As for previous calculations, cross sections are computed for collision energies from thresholds to 100 eV, and rate coefficients $\langle \sigma \varv \rangle$ then calculated and summed over all symmetries and cores. Final results are obtained for temperatures in the range $1000-20000$ K with steps of 1000 K, and the results for the rate coefficients are published electronically (see footnote on the first page for details).

\section{Results and discussion}
\begin{figure*}[h!]
    \centering
    \includegraphics[width=\linewidth]{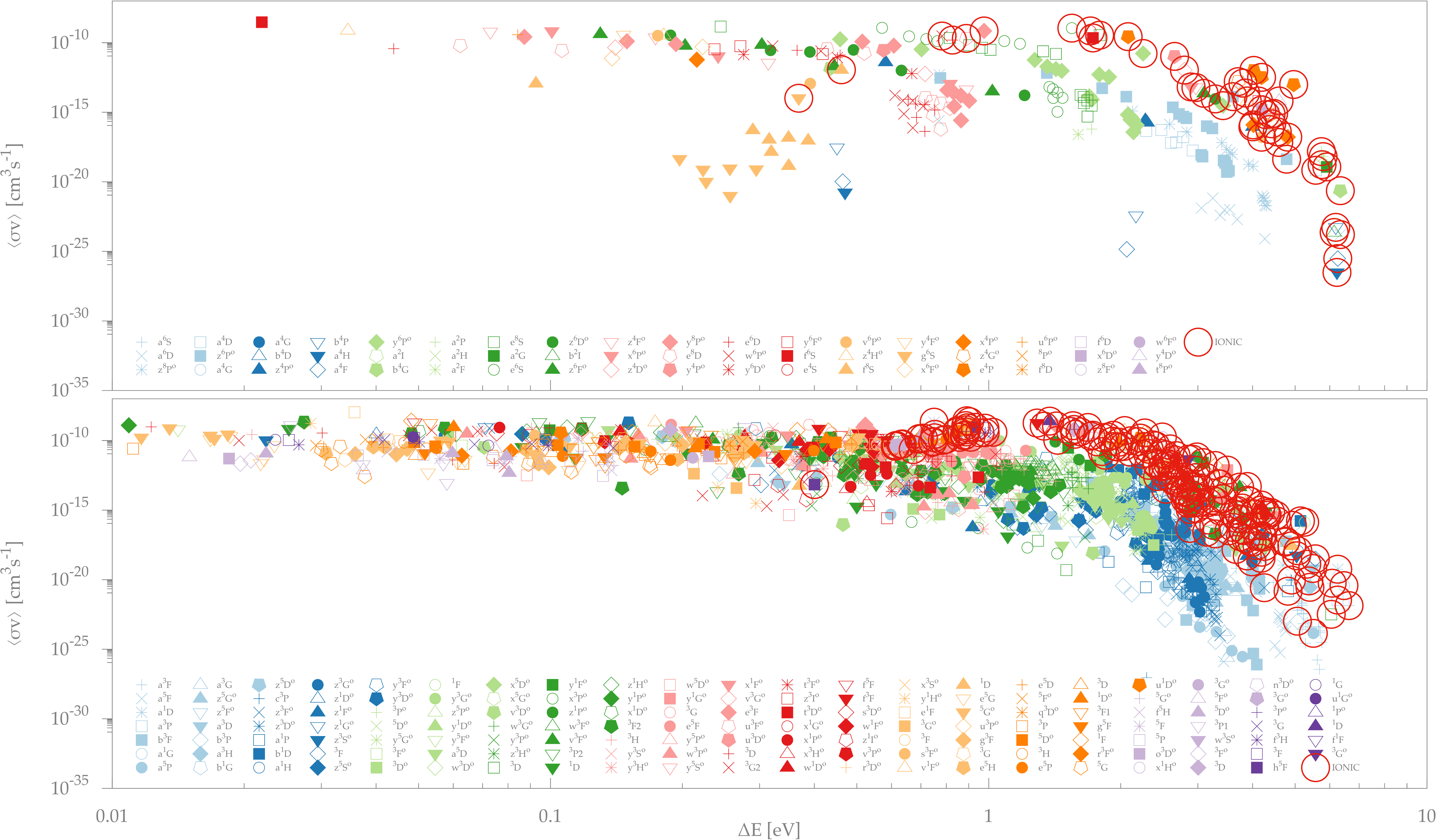}
    \caption{Rate coefficients $\langle\sigma v \rangle$ for Mn + H (\textbf{upper}) and Ti + H (\textbf{lower}) collision processes at 6000 K, plotted for various asymptotic energy differences between initial and final molecular states, $\Delta E$. The data shown is for endothermic processes only, i.e., excitation and ion-pair production. Legend labels are given for the initial state of Mn or Ti involved in the transition, and processes leading to a final state that is ionic are circled.}
    \label{fig:rates}
\end{figure*}

\begin{figure*}[ht!]
    \includegraphics[width=0.8\linewidth]{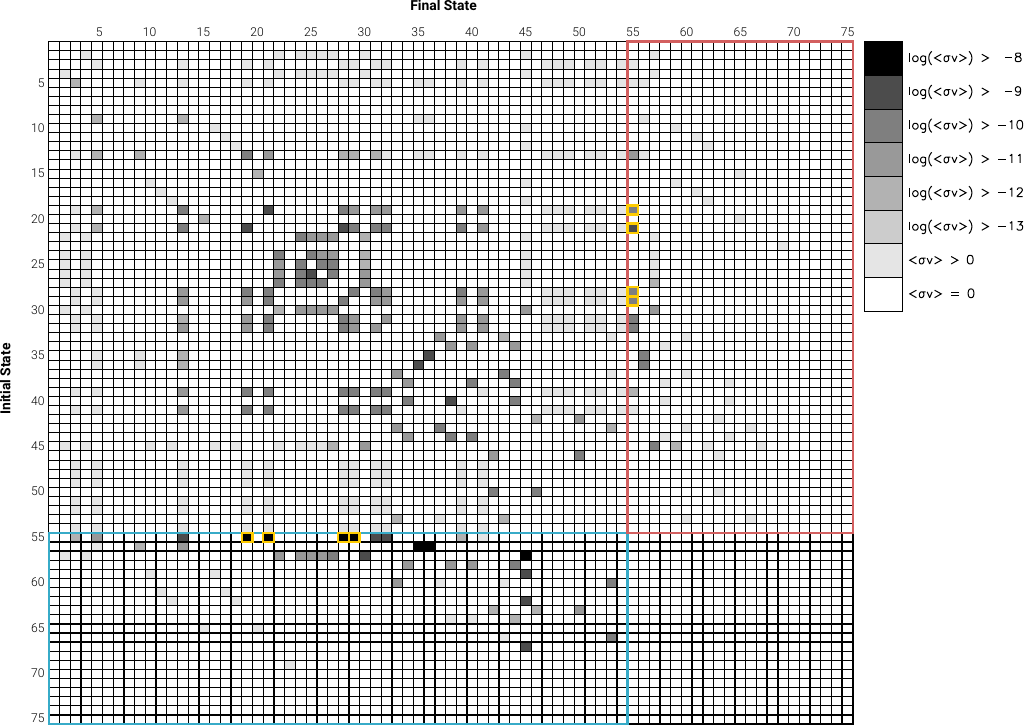}
    \\ \ \\
    \includegraphics[width=0.686\linewidth]{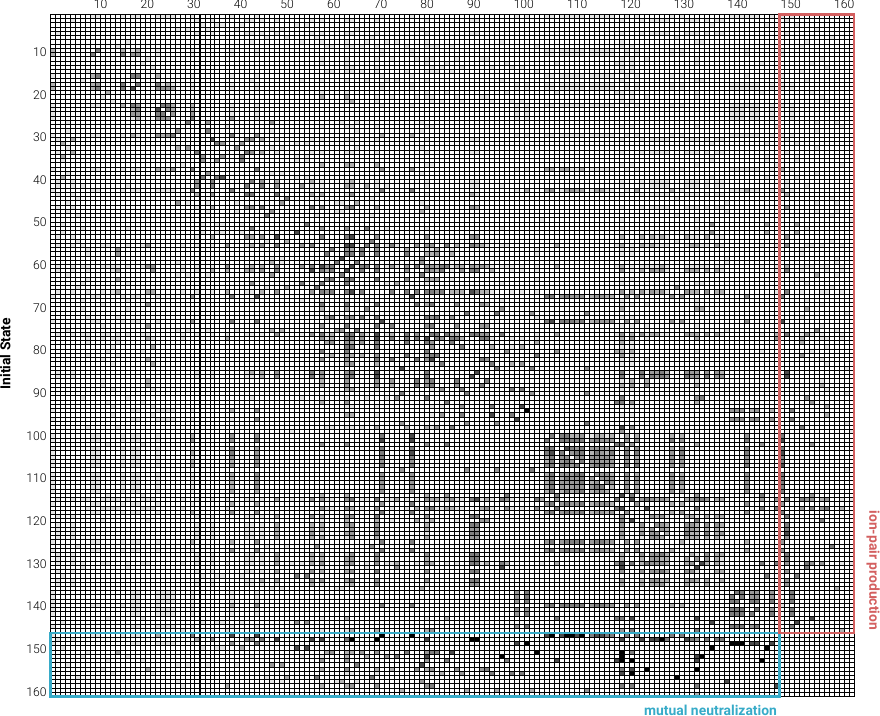}
    \caption{\label{fig:rates_grid}Heat map of the rate coefficient matrix $\langle \sigma \varv \rangle$ [cm$^3$~s$^{-1}$] for inelastic Mn + H / Mn$^+$ + H$^-$ collisions (\textbf{upper}) and Ti + H / Ti$^+$ + H$^-$ (\textbf{lower}) at $T = 6000$~K, calculated with the LCAO asymptotic model. The logarithms in the legend are to base 10. The matrix element indices correspond to those labeling the asymptotic states in Tabs. \ref{tab:states_Mn} and \ref{tab:states_Ti} for Mn and Ti, respectively, of which for Mn $1-54$ are covalent and $55-75$ ionic, while for Ti, $1-146$ are covalent and $147-161$ are ionic. The charge transfer processes, involving initial or final ionic states, are outlined with (colored) boxes, for ion-pair production in the upper right (red) and for mutual neutralization in the lower left (blue). For Mn, the four dominating mutual neutralization channels and their corresponding reverse ion-pair production channels (compared with full quantum calculations in Table \ref{tab:comparison_Mn}) are outlined in yellow.}
    \vspace{-2em}
\end{figure*}

 Figure~\ref{fig:pots_couplings} shows example potentials and couplings for a selected important core and symmetry of both Mn+H and Ti+H.  The plots both show the series of avoided crossings between ionic and covalent states; this is the mechanism for nonadiabatic transitions in the model.   Rate coefficients at 6000~K, obtained as the result of calculations for all possible cores and symmetries, are shown in Figs.~\ref{fig:rates} and~\ref{fig:rates_grid} in two different formats.  As has been seen in previous work, these plots show that charge transfer processes, ion-pair production $\mathrm{X}+\mathrm{H}\rightarrow\mathrm{X}^+ + \mathrm{H}^-,$
 and mutual naturalization
 $\mathrm{X}^+ + \mathrm{H}^-\rightarrow\mathrm{X}+\mathrm{H}$,
 provide the largest rate coefficients.  States roughly 1--2~eV below the ionic limit (corresponding to $\Delta E$ of the same magnitude, see Fig.~\ref{fig:rates}) lead to optimal crossing distances of around 14--27 a.u. for the low velocities of interest, resulting in the largest rate coefficients.

\begin{table*}
    \caption{Comparison of rate coefficients $\langle \sigma v \rangle$ [cm$^3$/s] for some important scattering channels with \cite{belyaev_atomic_2017}, labeled BV17, at $T~=~6000$~K. Column 1-4 labels the involved ionic or covalent asymptotic states, where the indices correspond to the states listed in Table \ref{tab:states_Mn}. Columns 5-6 and 8-9 list rate coefficients for various processes defined by the subheadings, and columns 7 and 10 labeled $\delta_\textrm{BV17}$ give the quotient of the rates from this work with the results from BV17. Only channels involving at least one rate coefficient larger than $10^{-9}$ cm$^3$/s in either this work or BV17 are included. All numbers are rounded off to two significant digits, and square brackets denote powers of ten (i.e., $ [X] \equiv 10^X$).
    }
    \label{tab:comparison_Mn}
    \centering
    \setlength\tabcolsep{0pt} 
    \begin{tabular*}{\textwidth}{c @{\extracolsep{\fill}} cccllllll}
    \toprule
    \textbf{Index} & \boldmath ${\alpha~^{2S+1}L}$   &  \bf Index & \boldmath $\alpha~^{2S+1}L$ & \bf This work     & \bf BV17 &  \boldmath $\delta_\textrm{BV17}$ & \bf This work &  \bf BV17 & \boldmath $\delta_\textrm{BV17}$ \\ \midrule
\multicolumn{2}{c}{Ionic} & \multicolumn{2}{c}{Covalent} & \multicolumn{3}{l}{Mutual Neutralization} & \multicolumn{3}{l}{Ion-Pair Production} \\
\cmidrule{1-2} \cmidrule{3-4} \cmidrule{5-7} \cmidrule{8-10}
55 & a $^7$S & 13 & y $^6$P$^o$ & $6.7[-9]$ & $6.5[-9]$ & $ 1.0$     & $1.7[-11]$ & $1.6[-11]$ & $ 1.0$  \\
   &         & 19 & e $^8$S     & $2.5[-8]$ & $2.0[-8]$ & $ 1.3$     & $3.5[-10]$ & $2.7[-10]$ & $ 1.3$  \\
   &         & 21 & e $^6$S     & $3.8[-8]$ & $4.4[-8]$ & $ 0.87$    & $1.1[-9] $ & $1.3[-9] $ & $ 0.88$ \\
   &         & 28 & y $^8$P$^o$ & $3.2[-8]$ & $2.7[-8]$ & $ 1.2$     & $7.0[-10]$ & $6.0[-10]$ & $ 1.2$  \\
   &         & 29 & e $^8$D     & $1.3[-8]$ & $1.6[-8]$ & $ 0.82$    & $2.0[-10]$ & $2.5[-10]$ & $ 0.82$ \\
   &         & 31 & e $^6$D     & $8.8[-9]$ & $9.4[-9]$ & $ 0.93$    & $1.6[-10]$ & $2.2[-10]$ & $ 0.70$ \\
   &         & 32 & w $^6$P$^o$ & $7.0[-9]$ & $5.9[-9]$ & $ 1.2$     & $3.0[-10]$ & $2.5[-10]$ & $ 1.2$  \\
     \midrule
\multicolumn{4}{c}{Covalent} & \multicolumn{3}{l}{De-excitation} & \multicolumn{3}{l}{Excitation} \\
\cmidrule{1-4} \cmidrule{5-7} \cmidrule{8-10}
13 & y $^6$P$^o$ & 19 & e $^8$S     & $9.6[-10]$ & $2.0[-9]$  & $0.48$ & $1.8[-10]$ & $3.7[-10]$ & $ 0.48$  \\
19 & e $^8$S     & 21 & e $^6$S     & $3.1[-9]$  & $4.8[-9]$  & $0.65$ & $1.5[-9]$  & $2.2[-9]$  & $ 0.65$  \\
21 & e $^6$S     & 28 & y $^8$P$^o$ & $8.9[-10]$ & $7.0[-10]$ & $1.3$ & $1.2[-9]$  & $9.3[-10]$ & $ 1.3$   \\
&&&&&&&& \\ \bottomrule
    \end{tabular*}
    \label{tab:my_label}
\end{table*}

In the case of Mn, the dominant channels for mutual neutralization from Mn$^+(a~^7\mathrm{S})$+H$^-$ are found to be into the Mn states $e~^6$S (index 21), $y~^8$P$^o$ (28), $e~^8$S (19) and $e~^8$D (21), all having rate coefficients larger than $10^{-8}$~cm$^3/$s at 6000~K.  In addition, mutual neutralization from the excited ionic state Mn$^+(a~^5\mathrm{S})$+H$^-$ into $f~^6$S (index 35), and $e~^4$S (36) are also larger than $10^{-8}$~cm$^3/$s. Furthermore, mutual neutralization from the excited ionic state Mn$^+(a~^5\mathrm{D})$+H$^-$ into $e~^4$P (index 45) is predicted to exceed $10^{-8}$~cm$^3/$s.

For Ti, the situation is much more complicated with large mutual neutralization rates into a significant number of states.  Generally, from the ground state ionic channel Ti$^+(a~^4\mathrm{F})$+H$^-$ mutual neutralization occurs into a cluster of states around index 110 with about 5.2~eV excitation, meaning about 1 eV below the ionic limit.  Similar to the trend seen in Mn, excited ionic states Ti$^+(b~^4\mathrm{F})$+H$^-$ and Ti$^+(a~^2\mathrm{F})$+H$^-$ lead to large mutual neutralization rates in slightly more excited states. It may also be noted that large mutual neutralization rates into states $^3$F1 and $^1$F (indexes 115 and 117) are seen from a number of excited ionic channels.

Excitation and de-excitation rates are typically smaller than charge transfer, as they must involve transitions at two ionic crossings, but can still reach reasonably large values of order $~10^{-10}$~cm$^3/$s, and even $~10^{-8}$~cm$^3/$s for individual transitions.  Similar to what was seen in earlier studies of other atoms, such moderate to large rate coefficients for excitation and de-excitation are seen for near-lying states, especially among clusters of near-lying states.   In particular, in Mn there is a cluster of states with excitation energies of around 5.4 to 5.7~eV ($z~^6\mathrm{F}^o$, $z~^4\mathrm{F}^o$, $x~^6\mathrm{P}^o$, $z~^4\mathrm{D}^o$, indexes 24 to 27), with moderate rate coefficients.  In addition,\ some isolated pairs of near-lying states show rate coefficients of similar magnitude, for example $e~^8$S~--~$e~^6$S (indexes 19 to 21) and $f~^6$S~--~$e~^4$S (35 to 36), involving the same states as those with large mutual neutralization rates.
In Ti, the main cluster of moderately coupled states is seen around state 110, though there are other smaller clusters and isolated pairs of near-lying states also showing significant rate coefficients.

Comparing the calculated rate coefficients $\langle \sigma v \rangle$ for Mn with the results of BV17 gives both a check on our calculations, as well as an indication of the uncertainties involved in these types of asymptotic model calculations.  The main differences between our calculations and those of BV17 are: the method for calculation of the couplings (LCAO vs. semiempirical method); and the number of ionic states included (BV17 include the dominant ground ionic state, while here some excited ionic states are included).  In Table \ref{tab:comparison_Mn}, we present the four dominating scattering channels at $T = 6000$ K, namely the mutual neutralization of Mn$^+$(3d$^5$\,4s\,$^7$S) + H$^-$(1s$^2$) into Mn(3d$^5$\,4s($^7$S)\,$nl\,^{2\textrm{S}+1}$L) + H(1s) with $nl$ = 5s (index 19, 21), 5p (28), and 4d (29), as well as the corresponding reversed ion-pair production channels. The table also includes a comparison of moderately large mutual neutralization rate coefficients $> 10^{-9}$ cm$^3/$s (index 13, 31 and 32). Finally, the table lists the excitation and de-excitation processes with rate coefficients in the same moderately large range.

Generally, the results are in quite good agreement, with mutual neutralization rates agreeing within 30~\%, and de-excitation rates roughly within a factor of 2.  These differences are indicative of the relative uncertainties involved, but without doubt underestimate the true uncertainties. Previous work comparing asymptotic calculations with full quantum results in simple systems, as well as comparison of results of different asymptotic models, has led to the general conclusion that uncertainties are perhaps around a factor of 2 for the very largest rates, roughly one order of magnitude for large to moderate rates, and becoming larger for smaller rates (see, e.g., B16).  The uncertainties in the largest rates are mostly expected to stem from the shortcomings of the model calculations of the interactions and from the simplified treatment of the dynamics.  For the transitions with moderate to small rates, the assumption that the ionic crossing mechanism is dominant is questionable, and could lead to significant underestimations of the rates \citep[e.g.,][]{BelyaevInelasticOxygen2019}.

\begin{acknowledgements}
This work received financial support from the Swedish Research Council (2016-03765) and the project grants ``The New Milky Way'' (2013.0052) and ``Probing charge- and mass- transfer reactions on the atomic level'' (2018.0028) from the Knut and Alice Wallenberg Foundation.
\end{acknowledgements}

\bibliographystyle{aa}
\bibliography{refs,new_refs}

\end{document}